\documentclass{elsart}
\usepackage{graphics}

\def\BA{\begin{eqnarray}}
\def\BE{\begin{equation}}
\def\EA{\end{eqnarray}}
\def\EE{\end{equation}}

\def\eps{\varepsilon}

\def\gtsim{\lower-0.45ex\hbox{$>$}\kern-0.77em\lower0.55ex\hbox{$\sim$}}
\def\ltsim{\lower-0.45ex\hbox{$<$}\kern-0.77em\lower0.55ex\hbox{$\sim$}}

\begin{document}
\begin{frontmatter}
\hyphenation{Coul-omb ei-gen-val-ue ei-gen-func-tion Ha-mil-to-ni-an
  trans-ver-sal mo-men-tum re-nor-ma-li-zed mas-ses sym-me-tri-za-tion
  dis-cre-ti-za-tion dia-go-na-li-za-tion in-ter-val pro-ba-bi-li-ty
  ha-dro-nic he-li-ci-ty Yu-ka-wa con-si-de-ra-tions spec-tra
  spec-trum cor-res-pond-ing-ly}
\title{Wilson Lines Near the Light Cone}
\author{ Hans J.~Pirner$^{\rm a,b}$}
  \address{$^{\rm a}$ Institut f\"ur Theoretische Physik der
Universit\"at, Philosophenweg 19, D-69120 Heidelberg, Germany\\
    $^{\rm b}$
      Max-Planck-Institut f\"ur Kernphysik,
      Postfach 103980, D-69029 Heidelberg, Germany }
\date{15 August 2000}
\begin{abstract}
We study the dynamics of  Wilson lines near the light cone in QCD.
Lattice simulations of the near light cone Hamiltonian in $SU(2)$ 
show that the correlation mass at
strong 
  coupling vanishes at intermediate coupling, which signals a continuum
  transition on the light cone. We point out the possible relevance of
this mass for high energy scattering.  
\end{abstract}
\end{frontmatter}

\section{Introduction}
  Many empirical data point towards a nontrivial vacuum structure in
QCD. 
The question arises what is the effect of vacuum 
condensates
on the dynamics near the light  cone.
High energy reactions have a very elegant formulation in 
light cone coordinates which present an enormous simplification,
by reducing the number of relevant diagrams in perturbation theory.
In recent years the extension of light cone theory to a calculation of 
wave functions and bound state spectra, i.e. soft physics has been
presented in previous light cone conferences and reviews \cite{Brodsky}. 
In light cone theory a trivial vacuum
is the starting basis. Naively spoken, with only positive momentum fractions
complicated many particle configurations cannot be contained in the 
vacuum.
In this theory on the light cone with a trivial vacuum, the 
difficult object is not the vacuum, but the Hamiltonian itself which
may contain terms beyond the naive Lagrangian 
which result from the elimination of
one half of the Hilbert space, namely the negative energy states.

We propose an approach \cite{NPFV,HJP} which starts near the light
cone, preserving the nonperturbative  features of QCD, but aims 
at the derivation of an effective light cone Hamiltonian 
in a controlled limiting procedure. 
The Cauchy problem is well defined in 
near light cone coordinates, since the initial data are given on a 
space-like surface. Such a  formulation avoids the solution of 
constrained equations, which on the quantum field theoretic 
level may be very complicated. The problem of a nontrivial 
vacuum appears in a solvable form related to the transverse
dynamics. This is physically very appealing, since in high 
energy reactions the incoming particles propagate near the
light cone and interact mainly through particle exchanges
with transverse momenta.

Near light cone QCD has a nontrivial vacuum which 
cannot be neglected even in the light cone limit. 
Generally, if the zero mode theory has massive excitations,
then these masses diverge like $\frac{1}{\eta}$
in the light cone limit ${\eta \rightarrow 0}$.
Massless excitations in the 
zero mode theory, however,  do not decouple in the
light cone limit. Genuine nonperturbative techniques must be
used to investigate the behaviour of this limit. In principle
the additional parameter which labels the coordinate system
can be chosen arbitrarily. The zero mode
Hamiltonian depends on an effective coupling constant containing
this parameter and evolves towards an infrared fixed point. 
We choose the following near light cone coordinates which
smoothly interpolate between the Lorentz and light front
coordinates:
\BA
  x^t ~=~ x^{+} &=& \frac1{\sqrt2} 
       \left\{ \left(1 + \frac{\eta^2}{2} \right)
       x^{0} + \left(1 - \frac{\eta^2}{2} \right) x^{3} 
       \right\} ~,\nonumber \\
          x^{-} &=& \frac1{\sqrt2} \left( x^{0}-x^{3} \right)~.
\label{Coor}
\EA
The transverse coordinates $x^1$, $x^2$ are unchanged;
$x^t = x^{+}$ is the new time coordinate, $x^{-}$ is a 
spatial coordinate. As finite quantization volume we will 
take a torus and its extension in spatial ``-'' direction, 
as well as in ``1, 2'' 
direction is $L$. The scalar product of two 4-vectors $x$ and $y$
is given with $\vec{x}_\bot \vec{y}_\bot = x^1 y^1 + x^2 y^2$ as
\BA
  x_\mu y^\mu & = & x^- y^+ + x^+ y^- - \eta^2 x^- y^-
                - \vec{x}_\bot \vec{y}_\bot \nonumber \\
              & = & x_- y_+ + x_+ y_- + \eta^2 x_+ y_+
                - \vec{x}_\bot \vec{y}_\bot ~.
\label{scalpr}
\EA
Obviously, the light cone is approached as the parameter $\eta$ 
goes to zero.
The gauge fixing procedure in the modified light-cone gauge 
$\partial_-A_- = 0$ allows  zero modes dependent on the transverse 
coordinates only. These zero mode fields carry zero linear momentum $p_-$
in near light cone coordinates, but finite amount of $p_0+p_3$.
They correspond to ``wee" or low $x$ - partons in the language of 
 Feynman. In color $SU(2)$ the zero-mode fields $a^3_-(x_\perp)$
can be chosen color diagonal proportional to $\tau^3$. 
The use of an axial gauge is very natural for the light-cone 
Hamiltonian even more so than in the equal-time Hamiltonian.
The asymmetry of the background zero mode naturally coincides 
with the asymmetry of the space coordinates on the light cone. 
The zero-mode fields describe disorder fields. Depending on the
effective coupling the zero mode transverse system will be in the
massive or massless phase. 
Our main objective has been to derive the precise relation
between the nearness to the light cone $\eta$ and the transverse resolution 
$a$. In principle these are two independent parameters. 
A considerable  simplification can be achieved if the limit 
towards the light cone is synchronized with the continuum limit.

\section{Near Light Cone QCD Hamiltonian} 

In ref. \cite{NPFV} we have derived the near light cone Hamiltonian.
We refer to this paper for further details.
The light cone Hamiltonian on the 
finite light like $x^-$ interval of length L has Wilson line or 
Polyakov operators
similarly to QCD formulated on a finite interval in imaginary time
at finite temperature.
\BE
  P(\vec{x}_\bot) = \frac12 \, \mbox{tr~P} 
  \exp\left(ig \int\limits_0^L dx^-A_-(\vec{x}_\bot,x^-)\tau^3/2\right)\,.
\label{Polyakov}
\EE

The underlying Hamiltonian governing the dynamics of these
Polyakov operators is a part of the full Hamiltonian $H$:
\BE
  H = \int d^3\!x \, H(\vec{x})\,,
\label{Ham}
\EE
with
\vskip-7mm
\BA
   H & = & \mbox{tr} 
    \left[ \partial_1 A_2 - \partial_2 A_1 - ig [A_1, A_2] \right]^2 
  + \frac1{\eta^2} \,\mbox{tr}
    \left[ \vec\Pi_\bot -
       \left( \partial_ - \vec{A}_\bot - ig [a_-,\vec{A}_\bot] \right)
    \right]^2
    \nonumber \\
  & + &  \frac1{\eta^2} \,\mbox{tr}
    \left[ \frac1L \vec{e}_\bot^{\;3} - \nabla_\bot a_- \right]^2 
    + \frac1{2 L^2} p^{-\dagger}(\vec{x}_\bot) p^{-}(\vec{x}_\bot)
    \nonumber \\
  & + & \frac1{L^{2}} \int\limits_0^L dz^- \int\limits_0^L dy^-
    \sum_{p,q,n}\!' 
      \frac{G_{\bot qp}(\vec{x}_\bot, z^-)%
            G_{\bot pq}(\vec{x}_\bot, y^-)}{\left[
      \frac{2\pi n}{L} + g(a_{-,q}(\vec{x}_\bot)
         - a_{-,p}(\vec{x}_\bot)) \right]^2 }
      e^{ 2\pi i n (z^- - y^-)/L} ~.
\label{Hamil}
\EA
The prime indicates that the summation is restricted to $n \neq 0$ 
if $p = q$. The operator $G_{\perp}$ gives the right hand side of
Gauss's law:
\BE
  G_{\perp}\left(\vec{x}\right) =
     \vec{\nabla}_\perp \vec\Pi_\perp \left(\vec{x}\right)
   + gf^{ab3} \frac{\lambda^a}2 \vec{A}\,^{b}_\perp
     \left(\vec{x}\right) \left(\vec\Pi\,^{3}_\perp
     \left(\vec{x}\right)
   - \frac1L \vec{e}\,^{3}_\bot \left(\vec{x}_\perp\right)\right) 
   + g\rho_m(\vec{x}) ,
\label{Gop}
\EE
with $\rho_m$ the matter density. The above Hamiltonian shows rather 
clearly that a naive limiting procedure $\eta \rightarrow 0$ does not
work. There are severe divergencies in this limit. The diverging
terms reappear in the usual light cone Hamiltonian as constraint equations
which are extremely difficult to solve on the quantum level in $3+1$
dimensions. 
The zero mode part of the full Hamiltonian is coupled to the
three-dimensional modes in the rest of the Hamiltonian. In the following
we will concentrate on universal properties of the zero mode Hamiltonian
which will survive the renormalization of the $(2+1)$ transverse dynamics
due to the coupling to the $(3+1)$  dimensional rest.
The zero mode
Hamiltonian contains  the Jacobian $J(a_-)$ which 
takes into account the Haar measure of $SU(2)$
$J\left(a_-(\vec{x}_{\bot})\right) = 
   \sin^2 \left( \frac{gL}{2}a_-(\vec{x}_{\bot})\right)$.   
It stems from the gauge fixing procedure, effectively introducing 
curvilinear coordinates. It also appears in the functional
integration volume element for calculating matrix elements.
It is convenient to introduce dimensionless variables 
\BE
  \varphi(\vec{x}_\bot) = \frac{gL a_-(\vec{x}_\bot)}{2} \,,
\EE
which  vary in a compact domain $0 \leq \varphi \leq \pi$. We 
regularize the above Hamiltonian $h_{\rm red}$ by introducing
a lattice constant $a$ on the lattice in the transverse directions. 
Next we appeal to the 
physics of the infinite momentum frame and factorize the reduced
true energy from the Lorentz boost factor $\gamma=\sqrt2/\eta$
and the cut-off by defining $h_{\rm red}$
\BE
  h = \frac1{2\eta a} h_{\rm red}.
  \label{hdef}
\EE

For small lattice spacing we obtain the reduced Hamiltonian
\BE
  h_{\rm red} = \sum_{\vec{b}}\left\{ -g^2_{\rm eff}
    \frac1J \frac\delta{\delta\varphi(\vec b)}
          J \frac\delta{\delta\varphi(\vec b)}
   + \frac1{g^2_{\rm eff}} \sum_{\vec\eps}
    \left(\varphi(\vec{b})-\varphi(\vec{b}+\vec\eps\,)\right)^2
  \right\} ~.
\EE
with the effective coupling constant
\BE
 g^2_{\rm eff} = \frac{g^2L\eta}{4a}.
\EE
In the continuum limit of the transverse lattice theory we let
$a$ go to zero.
In ref. \cite {HJP} we have done a Finite Size Scaling (FSS) 
\cite{Binder} analysis obtaining 
a second order transition between a phase with massive
excitations at strong coupling and a phase with massless excitations in weak
coupling.  The critical coupling is calculated as  
$g^{*2}= 0.17 \pm 0.03$, which is, however, not a universal 
quantity and subject to renormalization from the $(3+1)$ 
dimensional modes. A calculation in the epsilon expansion \cite{LZ}
gives the zero of the $\beta$-function as an infrared stable fixed point. 
Therefore the  limit of large transverse and longitudinal dimensions 
$L$ is well defined. Using the running coupling constant  
\BE
  g^2_{\rm eff} = 
  g^{*2} + \left(\xi_0 g^{*2}\right)^{1/\nu}
  \left(\frac{a}{L}\right)^{1/\nu} \,,
\EE
the lattice constant over the correlation length $L$ approaches zero at the
critical coupling:
\BE
  \frac{a}{L} =
  \frac{1}{\xi_0 g^{*2}} \left(g^2_{\rm eff}-g^{*2}\right)^{\nu} \,,
\EE
with $\nu=0.56 \pm 0.05 $. If we infer in addition that the coupling to the
three-dimensional modes will produce the usual running gauge coupling 
$g^2$ of $SU(2)$ QCD:
$g^2 = \frac{g^{2}_0}{\log\left(L/a\right)}$, then
we can synchronize the approach to the light cone, i.e. the limit 
$\eta\rightarrow0$ with the continuum limit $a\rightarrow0$. The 
condition that the three-dimensional evolution of $g^2$ has to be
compatible with the two-dimensional evolution of $g^2_{\rm eff}$ 
yields that for $a\to0$ the light cone parameter $\eta$ approaches
zero as
\BE
  \eta(a) = \frac{4a}{L}\frac{g^{*2}}{g^2} \,.
\EE

\section {High Energy Scattering and Wilson Lines}

The dynamics of 
high energy reactions 
is characteristically different from a description of
a single hadron. The fast projectile partons propagate near
one light cone direction, whereas the target partons cut the
trajectories of the projectile partons coming from the
opposite light cone direction. Balitsky \cite {Bal} has coined the 
phrase that a shock wave is encountered by the fast partons 
piercing through the target. 
In a theory of total cross sections the nonperturbative 
infrared dynamics in the transverse plane is essential. 
The model of  the stochastic vacuum \cite {dosim,DiGia} 
allows to connect confinement
with high energy diffractive scattering. The underlying idea  is to 
approximate the projectile and target by two color dipoles.
The color and anticolor components of these dipoles acquire phase factors
running along the respective light cones. These phase factors are
path ordered exponentials, i.e. Wilson lines along the light cone. 
Polyakov variables near the light
cone enter the calculations of high energy cross sections
in refs.~\cite{DGKP}.
After a transformation to Minkowski space time nonvanishing 
correlators of the stochastic vacuum model arise  
between gauge field strengths $<F_{+i}F_{-j}>$
arising from the phases on opposite light cones.  
For these calculations correlations between 
Polyakov lines on lightlike lines along the projectile and target 
directions are important.

There are two inherent deficiencies of a naive translation of the
Gaussian stochastic model to the  light cone. Firstly, the correlation function
of Wilson lines along the same light cone, i.e. correlators
$<F_{+i}F_{+j}>$ vanish. Here the more carefully derived 
Hamiltonian in this paper can make an important contribution. It generates
a transverse confining interaction for particles propagating along the
same light cone. The correlations in transverse direction depend on the
nearness to the light cone. Secondly, the model of soft 
scattering does lead to an energy independent high energy cross section.
This is not so bad an approximation for large dipole-dipole scattering,
like proton-proton scattering where the energy dependence is weak 
$s^{0.08}$, but
for the scattering of small dipoles on the proton HERA experiments show 
a very strong energy dependence $s^{0.35}$. The weak energy dependence
can be interpreted as a weakly increasing cloud of very soft 
sea partons in the proton which with increasing energy produces 
a growing total cross section.
The increasing cross section of small dipoles 
may have hard gluons as its origin.  These arise from the radiative 
corrections to the correlation function of the Wilson line operators 
along the same light cone under evolution towards the light cone,
c.f. ref.
\cite {Bal}. This equation contains no external mass scale.  Here 
the above treatment of the near light cone nonperturbative 
dynamics may come in to give such a mass scale which limits the
diffusion into large transverse distances under evolution.


\end{document}